\documentclass[superscriptaddress,nobibnotes,amsmath,amssymb,notitlepage,oneocolumn,prl,longbibliography]{revtex4-1}

\usepackage{bm,mathptmx,braket}

\usepackage{graphicx,color,hyperref}
\usepackage{bm}
\usepackage[caption=false]{subfig}
\hypersetup{colorlinks=true, linkcolor=blue, citecolor=blue, urlcolor=blue} 

\graphicspath{{./img/}}

\begin{document}

\title{Generalized GMP Algebra for Three-Dimensional Quantum Hall Fluids of Extended Objects}

\author{Giandomenico Palumbo}
\affiliation{Department of Physics, University of Coimbra, Rua Larga, 3004-516 Coimbra, Portugal}
\email{giandomenico.palumbo@gmail.com}

\date{\today}

\begin{abstract}
We develop a geometric framework for three-dimensional quantum Hall fluids of extended objects (quasi-strings) in the presence of a strong three-form background field associated with a bundle gerbe. In the strong-field regime, fast internal dynamics is frozen and the low-energy kinematics is governed by generalized guiding-center variables consisting of vectorial and tensorial coordinates. We show that these guiding-center variables obey a noncommutative geometry giving rise to a three-dimensional generalization of the Girvin–MacDonald–Platzman (GMP) algebra for projected density operators. 
Moreover, we relate this algebra to the canonical quantization of a topological BF+BB theory whose level is identified with the Dixmier–Douady invariant.
Our results clarify the structure of incompressible quantum Hall–type phases and their geometric and topological features in three spatial dimensions.
\end{abstract}

\maketitle

	\section{Introduction}

The quantum Hall effect provides a paradigmatic example of how strong magnetic
fields fundamentally reorganize the kinematics of quantum matter.
In two dimensions, projection onto the lowest Landau level leads to a
noncommutative geometry of guiding-center coordinates \cite{Haldane2011}, whose algebraic
structure underlies the Girvin-MacDonald-Platzman (GMP) algebra of projected
densities and the emergence of incompressible quantum fluids \cite{GMP, Cappelli1993, Szabo, Palumbo2023}.
Extending these ideas to three spatial dimensions has proven challenging.
For point particles and band models, an uniform (Abelian) magnetic field does not lead to a fully three-dimensional Landau-level
quantization. Higher-dimensional generalizations of the GMP algebra introduce, either some vortex-density operators in addition to the usual charge density operators \cite{Tiwari} in time-reversal-invariant topological fluids or nonassociative structures for charge density operators via Nambu brackets \cite{Nambu,Neupert, Estienne,Chu2} whose
quantization remains unclear \cite{Szabo2019}.
A key observation is that higher-form gauge fields couple naturally to
\emph{extended objects}, suggesting that a consistent three-dimensional
generalization of quantum Hall physics should involve degrees of freedom beyond
point particles.\\
In this work, we develop a continuum framework for three-dimensional incompressible quantum
Hall fluids of closed quasi-strings in the presence of a
strong three-form background field $H$, namely the curvature tensor of a Kalb-Ramond gauge field \cite{Kalb} associated to bundle gerbes \cite{Murray1996,Sati}.
We show that, in direct analogy with the two-dimensional case, the strong-field
limit freezes fast internal dynamics, here corresponding to rapid rotations of
oriented area elements, while leaving slow collective degrees of freedom (the oriented area swept by the string, analogous to angular momentum for point particles).
The resulting low-energy kinematics is governed by generalized guiding-center
variables consisting of vectorial coordinates $x^i$ and antisymmetric tensorial
coordinates $X^{ij}$ \cite{Palumbo2025, Givens}.
The strong three-form
field induces second-class constraints that reduce the phase space in a manner
analogous to lowest-Landau-level projection in two dimensions \cite{Luo,Du1}.
This \emph{projection} eliminates fast internal modes and defines a
reduced guiding-center sector whose quantization, implemented via Dirac
brackets \cite{Henneaux1992}, yields a universal noncommutative algebra,
which defines a tensorial noncommutative geometry that generalizes the
guiding-center structure of the two-dimensional quantum Hall fluids.
We then construct projected density operators and
show that they satisfy a novel kind of sine algebra \cite{FAIRLIE1989}, namely a generalized GMP algebra.
In the long-wavelength limit, this algebra reduces to the Lie algebra of
volume-preserving diffeomorphisms, identifying incompressibility as the
organizing principle of the three-dimensional quantum Hall fluid.
These results are conceptually complementary to the noncommutative strings arising at the boundary of open membranes in strong three-form backgrounds, as discussed in the M-theory literature \cite{Bergshoeff,Kawamoto,Saemann2012}.
Moreover, we demonstrate that a bulk topological quantum field theory, known as BF+BB theory \cite{Baez,Palumbo2025} and characterized by a dimensionless level identified with the Dixmier–Douady invariant \cite{Murray1996}, reproduces the three-dimensional guiding-center algebra upon canonical quantization. 
Our work establishes a direct analogy between two- and
three-dimensional quantum Hall physics, 
providing a unified and physically grounded description of incompressible three-dimensional fluids
of extended objects and opens the way to
systematic studies of their collective dynamics and responses.

\section{Generalized guiding centers and GMP algebra}

The GMP algebra plays a central role in the theory of the two-dimensional quantum Hall effect (QHE): it encodes the noncommutative geometry of guiding, namely center coordinates induced by a strong and uniform magnetic field and controls the structure of density fluctuations, collective modes, and fractionalized phases. In its most basic form, the GMP algebra arises from the projection of particle coordinates onto a Landau level, where the canonical coordinates cease to commute and acquire a universal commutator proportional to the magnetic length squared. The projected density operators then satisfy a closed ``sine algebra'' \cite{FAIRLIE1989}, whose long-wavelength limit reproduces the area-preserving diffeomorphism algebra of an incompressible quantum fluid.
The purpose of this paper is to develop and physically justify a three-dimensional analogue of the GMP algebra appropriate for
\emph{quantum Hall fluids in the presence of a strong background three-form field} $H$ in three spatial dimensions. This situation may be viewed as a higher-form generalization of the QHE: while in 2D the magnetic field is a closed two-form $F=\mathrm{d}A$ coupling to point-like particles, in 3D the natural background field is a closed three-form $H=\mathrm{d}B$ (a Kalb-Ramond field strength) \cite{Kalb}, which couples to extended objects and induces a fundamentally different geometric structure. Our main claim is that, in this setting, the correct kinematic variables are not only vectorial coordinates $x^i$, but also
\emph{antisymmetric tensorial coordinates} $X^{ij}=-X^{ji}$ \cite{Palumbo2025, Givens} and that their mutual noncommutativity gives rise to a well-defined three-dimensional GMP-type algebra.
In two spatial dimensions, the appearance of noncommutative coordinates is most transparently understood semiclassically. For a charged particle in a strong, uniform magnetic field $\mathcal{B}$, the cyclotron motion freezes and the guiding-center coordinates satisfy
\begin{equation}
	[x^i,x^j]= i \ell_B^2,\epsilon^{ij},
\end{equation}
with $\ell_B=1/\sqrt{e \mathcal{B}}$ the magnetic length (where we have set $\hbar=1$). Projected density operators $\rho(\mathbf{q})=e^{i q_i x^i}$ (with $\mathbf{q}$ the momenta) then obey the GMP algebra \cite{GMP}
\begin{equation}
	[\rho(\mathbf{q}), \rho(\mathbf{q'})]
	=
	-2i
	\sin\!\left(
	\frac{\ell_B^2}{2}
	\, q_i q_{j} \epsilon^{ij}
	\right)
	\rho(\mathbf{q+q'}),
	\label{eq:2dgmp}
\end{equation}
whose structure constants are fixed entirely by the symplectic form defined by $\mathcal{B}$.
In three spatial dimensions, a closed two-form is no longer the natural top-degree background field. Instead, the analogue of a ``magnetic field'' is given by
\begin{equation}
	H = \frac{1}{3!} H_{ijk}\, \mathrm{d}x^i\wedge\mathrm{d}x^j\wedge\mathrm{d}x^k ,
	\hspace{0.3cm} H_{ijk}=\partial_i B_{jk}+\partial_j B_{ki}+\partial_k B_{ij},
\end{equation}
which defines a volume form rather than an area form. Such three-form fields and corresponding Kalb-Ramond fields $B$ appear naturally in several contexts: string and membrane theory \cite{Kalb}, gravity \cite{Freidel2012}, spin foams \cite{Baez2000}, interacting topological insulators \cite{Cirio,Fradkin}, fermion-fermion dualities \cite{Palumbo2020,Grushin}, non-Abelian higher-rank symmetries \cite{Du2022} and hydrodynamic descriptions of incompressible polar fluids \cite{Tiwari}. Crucially, a three-form does not define a symplectic structure on ordinary coordinates $x^i$; therefore a direct generalization of the 2D associative guiding-center algebra is impossible if one insists on point-like coordinates alone.\\
In this work, we consider a three-dimensional quantum fluid whose fundamental constituents are
closed quasi-strings.
In fact, it is well known that such extended objects are the natural degrees of freedom that couple minimally
to the two-form gauge potential $B$.
This setup is nothing but the higher-form analogue of a charged quantum fluid coupled to an external magnetic field in two dimensions.
The fluid is assumed to be:
\begin{itemize}
	\item homogeneous, isotropic and incompressible at long wavelengths,
	\item subject to a strong and uniform three-form background field, i.e.
	$H_{ijk} = H \epsilon_{ijk}$,
	\item dominated by collective degrees of freedom rather than
	microscopic string excitations.
\end{itemize}
Our first goal is to identify the appropriate \emph{projected kinematic variables}
that survive in the strong-field limit. For this reason, let us consider the dynamics of a single closed string parametrized by $\sigma \in [0,2\pi)$ and embedded in
three-dimensional space via the worldsheet coordinates $X^i(\tau,\sigma)$, such that its action is given by \cite{Polchinski1998}
\begin{equation}
	S = S_{\text{kin}} + q \int_{\Sigma_2} B=  \frac{m}{2} \int d\tau d\sigma \, \partial_{\tau} X^i \partial_{\tau} X_i+ \frac{e}{2} \int d\tau d\sigma \, 
	B_{ij}(X)\, \partial_\alpha X^i \partial_\beta X^j \epsilon^{\alpha\beta},
	\label{eq:string-action}
\end{equation}
where $S_{\text{kin}}$ includes tension and inertial terms, $e$ is the effective charge (notice, we use the same symbol of the electric charge, although it can assume a different value) and $\alpha, \beta = \{\tau, \sigma\}$. Here, 
$X^i(\tau,\sigma)$ embeds the string worldsheet $\Sigma_2$ into space  and the second term in the action
is the Kalb-Ramond coupling.
In a uniform three-form background,
variation of \eqref{eq:string-action} with respect to $X^i$ yields the following force density
\begin{equation}
	f_i(\sigma) = e\, H \epsilon_{ijk}\,
	\partial_\tau X^j \partial_\sigma X^k,
\end{equation}
which is the direct analogue of the Lorentz force
$f_i = e\, \mathcal{B}\,\epsilon_{ij} \dot x^j$ acting on a point particle in two dimensions.
This force generates a
\emph{torque} proportional to the oriented area spanned by the string, i.e.
\begin{equation}
	\tau_{ij} \sim e\, H \epsilon_{ijk}\, \dot X^k ,
\end{equation}
which induces rigid rotational motion of the area element.
Thus, in a uniform three-form background, the dominant response of closed string is
\emph{rotation of the area it spans}, rather than circular motion of a point particle.
This rotational dynamics is the higher-form analogue of cyclotron motion.
Thus, in the regime
\begin{equation}
	|H| \gg \mathcal{E},
\end{equation}
where $\mathcal{E}$ denotes all other energy scales (string tension, interactions,
gradients), the dynamics separates into two classes:
\begin{itemize}
	\item \emph{Fast modes}: internal rotational motion of the string area, with
	characteristic frequency set by $|H|$.
	\item \emph{Slow modes}: collective motion of the string’s center and orientation.
\end{itemize}
This separation of scales is the direct analogue of the decomposition
of particle motion into fast cyclotron orbits and slow guiding-center motion in
the two-dimensional quantum Hall effect.
For our purposes, it is sufficient to calculate the canonical momentum conjugate  of $X^i(\sigma)$ from Eq.~\ref{eq:string-action} that reads
\begin{equation}
	\Pi_i(\sigma)
	= \frac{\delta L}{\delta \partial_{\tau} X^i}
	= m\, \partial_{\tau}X_i
	+ e B_{ij}(X) \partial_\sigma X^j,
	\label{eq:canonical-momentum}
\end{equation}
and take the strong-field limit $m \rightarrow 0$,
which is the precise analogue of the lowest-Landau-level projection in the
two-dimensional quantum Hall problem \cite{Luo,Du1}. In fact, in two dimensions, LLL projection occurs when $\omega_c \gg \mathcal{E}$ (i.e. with respect to all other energies in the system). Here, $|H| \gg \mathcal{E}$ plays the analogous role. The limit $m \to 0$ is taken after rescaling to dimensionless variables where $eH$ sets the dominant energy scale, analogous to setting the cyclotron energy as the reference scale in two dimensions.
In this limit, the momentum is no longer an independent variable but becomes
a constraint, i.e.
\begin{equation}
	\phi_i(\sigma)
	\equiv
	\Pi_i(\sigma)
	- e B_{ij}(X) \partial_\sigma X^j
	\approx 0.
	\label{eq:constraint}
\end{equation}
These are \emph{second-class constraints} in the sense of Dirac \cite{Henneaux1992}. The fundamental equal-time Poisson brackets are
\begin{equation}
	\{ X^i(\sigma), \Pi_j(\sigma') \}
	= \delta^i_j \delta(\sigma-\sigma'),
	\qquad
	\{ X^i, X^j \} = \{ \Pi_i, \Pi_j \} = 0.
\end{equation}
We now compute the Poisson brackets between constraints
\begin{align}
	\{ \phi_i(\sigma), \phi_j(\sigma') \}
	=
	- e\, H_{ijk}
	\, \partial_\sigma X^k
	\, \delta(\sigma-\sigma')-2e\,B_{ij}(X(\sigma))\,\partial_\sigma\delta(\sigma-\sigma'),
	\label{eq:constraint-algebra}
\end{align}
and remind that the Dirac brackets between two functionals $A,B$ are defined as \cite{Henneaux1992}
\begin{equation}
	\{A,B\}_D
	=
	\{A,B\}
	-
	\iint d\sigma d\sigma'
	\{A,\phi_i(\sigma)\}
	C^{-1}_{ij}(\sigma,\sigma')
	\{\phi_j(\sigma'),B\},
	\label{eq:dirac}
\end{equation}
where $C_{ij}(\sigma,\sigma') \equiv \{\phi_i(\sigma),\phi_j(\sigma')\}$.
From Eq. \eqref{eq:constraint-algebra}, one finds
\begin{equation}
	C^{-1}_{ij}(\sigma,\sigma')
	=
	\frac{1}{eH}
	\epsilon_{ijk}
	\frac{\partial_\sigma X^k}{|\partial_\sigma X|^2}
	\, \delta(\sigma-\sigma') ,
	\label{eq:Cinv}
\end{equation}
where the normalization is fixed by the oriented area of the quasi-string (see Appendix A for more details).
Notice that the second term on the lhs of Eq. (\ref{eq:constraint-algebra}) does not contribute to the Dirac brackets of the
collective observables considered here, since we restrict to a uniform
$H$-field and to reparametrization-invariant,
$\sigma$-integrated quantities. In this case, and in a linear gauge for
$B_{ij}$ compatible with constant $H=dB$, the term proportional to
$\partial_\sigma\delta$ reduces to a total derivative along the closed
string and gives no bulk contribution.
We can now define the collective (guiding-center) variables, given by
\begin{align}
	x^m &\equiv \frac{1}{2\pi} \oint d\sigma\, X^m(\sigma),
	\\
	X^{ij} &\equiv \frac{1}{2} \oint d\sigma
	\left(
	X^i \partial_\sigma X^j
	- X^j \partial_\sigma X^i
	\right).
\end{align}
The only non-vanishing Dirac bracket is (see Appendix A for more details)
\begin{align}
	\{ x^m, X^{ij} \}_D
	&=
	- \iint d\sigma d\sigma'
	\{x^m,\phi_k(\sigma)\}
	C^{-1}_{kl}(\sigma,\sigma')
	\{\phi_l(\sigma'),X^{ij}\},
\end{align}
and by employing
\begin{equation}
	\{x^m,\phi_k(\sigma)\}
	= \frac{1}{2\pi} \delta^m_k ,
\end{equation}
and
\begin{equation}
	\{\phi_l(\sigma),X^{ij}\}
	=
	\delta^i_l \partial_{\sigma} X^j
	- \delta^j_l \partial_{\sigma} X^i ,
\end{equation}
we obtain
\begin{align}
	\{x^m,X^{ij}\}_D
	&=
	-\frac{1}{2\pi}\oint d\sigma\;\frac{1}{eH}\,
	\epsilon^{m\ell p}\,\frac{\partial_{\sigma}X^{ p}(\sigma)}{|\partial_{\sigma}X(\sigma)|^{2}}\,
	\Big(\delta^i{}_\ell \partial_{\sigma}X^{j}(\sigma)-\delta^j{}_\ell \partial_{\sigma}X^{ i}(\sigma)\Big).
	\label{eq:intermediate_shape}
\end{align}
At this stage, the Dirac bracket in Eq.~(\ref{eq:intermediate_shape}) still retains explicit information about the local
tangent distribution along the loop through the combination
$\partial_\sigma X^i/|\partial_\sigma X|^2$.
As such, it depends on the detailed internal configuration of the quasi-string and is not yet
universal at the operator level.
However, in the strong-field regime $|H|\gg E$, the internal degrees of freedom of the closed
string undergo rapid rotational dynamics with a characteristic frequency set by $|H|$, while
collective guiding-center motion occurs at much lower energies. Upon quantization, this
separation of scales implies that the physical low-energy Hilbert space factorizes into a fast
internal sector and a slow collective sector, in direct analogy with the separation between
cyclotron and guiding-center degrees of freedom in the two-dimensional quantum Hall effect.
The relevant guiding-center algebra is therefore defined as an operator identity within the
projected internal ground-state sector.
Denoting by $|0_{\mathrm{int}}\rangle$ the rotationally invariant ground state of the frozen
internal modes, we define the effective guiding-center Dirac brackets through expectation
values of the corresponding operators in this sector. Rotational invariance of
$|0_{\mathrm{int}}\rangle$ then enforces
\begin{equation}
	\left\langle 0_{\mathrm{int}} \left|
	\frac{1}{2\pi}\oint d\sigma\,
	\frac{\partial_\sigma X^i(\sigma)\partial_\sigma X^j(\sigma)}
	{|\partial_\sigma X(\sigma)|^2}
	\right| 0_{\mathrm{int}} \right\rangle
	= \frac{1}{3}\,\delta^{ij},
	\label{eq:isotropic_ground_state}
\end{equation}
up to an overall normalization that can be absorbed into the definition of the tensorial
coordinates. This relation reflects the absence of any preferred direction in the internal
ground state and is the direct higher-form analogue of the isotropy of cyclotron motion in the
lowest Landau level (namely, we are assuming the existence of an SO(3)-invariant gapped ground state).

By employing Eq.~\eqref{eq:isotropic_ground_state}, and working for convenience in arc-length gauge
$|\partial_\sigma X|=1$, the projected Dirac bracket reduces to the universal guiding-center
form
\begin{equation}
	\left\langle 0_{\mathrm{int}} \left| \{x^m,X^{ij}\}_D \right| 0_{\mathrm{int}} \right\rangle
	= \ell_H^3\,\epsilon^{mij},
	\label{eq:dirac-result}
\end{equation}
where we have introduced the characteristic volume scale
\begin{equation}
	\ell_H^3 \equiv \frac{1}{eH}.
\end{equation}
Upon quantization, the projected Dirac brackets are promoted to commutators,
$\{A,B\}_D\rightarrow -i[A,B]$, yielding the universal generalized guiding-center algebra
\begin{equation}
	[x^m,X^{ij}] = i\ell_H^3\,\epsilon^{mij}, \qquad
	[x^i,x^j]=[X^{ij},X^{kl}]=0,
	\label{eq:final-gc-algebra}
\end{equation}
which holds as an operator identity within the low-energy projected Hilbert space.
Moreover, we emphasize that shape-dependent corrections correspond to excitations of the internal sector
are separated from the guiding-center physics by an energy gap of order $|H|$.
The above algebra is universal, i.e. it depends only on the background $H$ field
and the extended nature of the degrees of freedom, and not on microscopic details
of the quasi-string dynamics (it is important to bear in mind that this universality holds within the isotropic projected sector of frozen internal modes).
Furthermore, Eq. \eqref{eq:final-gc-algebra} expresses the fact that translations of the
guiding center and deformations of the oriented area do not commute.
In a quantum fluid composed of many such extended objects, density fluctuations
are naturally described by operators built from the projected variables.
We introduce the following generators
\begin{align}
	\rho(\mathbf{q}) = e^{i q_m x^m}, \hspace{0.3cm}
	\Upsilon(\mathbf{Q}) = e^{i Q_{ij} X^{ij}},
\end{align}
which probe, respectively, center-of-mass density and oriented-area density.Here, $q_m$ and $Q_{ij}$ are the conjugate momenta associated to $x^m$ and $X^{ij}$, respectively \cite{Palumbo2025}.
By employing the Baker--Campbell--Hausdorff formula and the algebra
\eqref{eq:final-gc-algebra} (see Appendix B for more details), one finds a closed commutation relation
\begin{equation}
	[\rho(\mathbf{q}), \Upsilon(\mathbf{Q})]
	=
	-2i
	\sin\!\left(
	\frac{\ell^3_H}{2}
	\, q_m Q_{ij} \epsilon^{mij}
	\right)
	e^{\,i q_m x^m + i Q_{ij} X^{ij}}.
	\label{eq:3dgmp}
\end{equation}
Equation \eqref{eq:3dgmp} is the three-dimensional generalization of the
GMP algebra and one of the main results of this paper.
In the long-wavelength limit, it reduces to the quantum algebra of
volume-preserving diffeomorphisms, reflecting the incompressibility of the
underlying quantum fluid as we will show in the next section.

\section{Volume-preserving diffeomorphisms and tensorial coordinates}
To describe diffeomorphisms in our framework, we work at the level of a \emph{fluid} of extended objects and in this section we
introduce coarse-grained, equal-time operator-valued densities. We therefore
promote the guiding-center variables to spatial fields
\begin{equation}
	x^i \;\to\; x^i(\mathbf{r}),
	\qquad
	X^{ij}\;\to\; X^{ij}(\mathbf{r}),
\end{equation}
and the local algebra consistent with the reduced guiding-center kinematics of Sec.~2 is
\begin{equation}
	\{x^m(\mathbf{r}),X^{ij}(\mathbf{r}')\}
	=\ell_H^3\,\epsilon^{mij}\,\delta^{(3)}(\mathbf{r}-\mathbf{r}'),
	\qquad
	\{x^i(\mathbf{r}),x^j(\mathbf{r}')\}=0,
	\qquad
	\{X^{ij}(\mathbf{r}),X^{kl}(\mathbf{r}')\}=0.
	\label{eq:local_guiding_center_algebra}
\end{equation}
The defining kinematic property of an incompressible quantum fluid is the absence of local density
fluctuations. In a continuum description, this constraint implies that the allowed long-wavelength
deformations of the fluid configuration space are \emph{volume-preserving diffeomorphisms} (VPDs). Infinitesimally, such transformations act on spatial coordinates as
\begin{equation}
	x^i \;\longrightarrow\; x^i + \delta x^i(\mathbf{r}),
	\label{eq:vpd_action}
\end{equation}
with the incompressibility condition
\begin{equation}
	\partial_i\,\delta x^i(\mathbf{r}) = 0.
	\label{eq:incompressibility_condition}
\end{equation}
In two spatial dimensions, the above constraint admits a scalar parametrization
$\delta x^{i} = \epsilon^{ij}\partial_{j}\phi$, leading to the familiar algebra of
area-preserving diffeomorphisms generated by a single function $\phi$. This structure
underlies the long-wavelength limit of the two-dimensional GMP algebra and provides the
hydrodynamic description of incompressible quantum Hall fluids \cite{Du2022}.
In three dimensions, however, no single scalar potential can generate all the VPD solutions. Instead, incompressible flows require at least two independent functions
or, equivalently, an antisymmetric tensor structure. This obstruction is the geometric origin
of the appearance of Nambu brackets \cite{Nambu} in classical treatments of three-dimensional incompressible
fluids \cite{Neupert}. As we now show, the tensorial guiding-center variables $(x^{i}, X^{ij})$ provide a
natural and fully associative realization of the VPD algebra without invoking Nambu
brackets.\\
Starting from the coarse-grained guiding-center phase space and corresponding Dirac brackets discussed previously,
we now construct the generators of infinitesimal incompressible deformations directly from
the tensorial coordinates $X^{ij}$. Let us consider the functional
\begin{equation}
	G[\Lambda] \;=\; \int d^{3}\mathbf{r} \, \Lambda_{ij}(\mathbf{r})\, X^{ij}(\mathbf{r}),
	\label{VPDgenerator}
\end{equation}
where $\Lambda_{ij}(\mathbf{r}) = -\Lambda_{ji}(\mathbf{r})$ is an arbitrary smooth closed two-form, namely $\Lambda_{ij}=\partial_i A_j-\partial_j A_i$ for some vector field $A_i$ locally.
Here the coordinate $\mathbf r$ labeling the guiding-center fields
$x^{i}(\mathbf r)$ and $X^{ij}(\mathbf r)$ plays the role of a label for the
collective degrees of freedom, rather than an active spatial coordinate.
Accordingly, the antisymmetric fields $\Lambda_{ij}(\mathbf r)$ entering
$G[\Lambda]$ are treated as parameters specifying infinitesimal incompressible
deformations, not as geometric two-forms on physical space.
Their variation under the transformation generated by $\delta x^{m}(\mathbf r)$
is therefore taken to be pure advection,
$\delta\Lambda_{ij}=\delta x^{m}\partial_{m}\Lambda_{ij}$,
rather than the full Lie derivative.
With this interpretation, the closure of the generator algebra derived below
follows directly from the guiding-center Dirac brackets.
In fact, by employing Eq.\eqref{eq:dirac-result}, the induced variation of the guiding-center coordinate $x^{m}$ is given by
\begin{align}
	\delta x^{m}
	&= \{x^{m}, G[\Lambda]\}_{D}
	= \int d^{3}\mathbf{r}\, \Lambda_{ij}(\mathbf{r})\,\{x^{m},X^{ij}\}_{D} \nonumber\\
	&= \frac{1}{eH}\,\epsilon^{mij}\,\Lambda_{ij}(\mathbf{r}),
	\label{VPDvariation}
\end{align}
and by taking the divergence of the above equation, we find
\begin{equation}
	\partial_{m}\delta x^{m}
	= \frac{1}{eH}\,\epsilon^{mij}\,\partial_{m}\Lambda_{ij} = 0.
\end{equation}
Therefore, the transformations generated by $G[\Lambda]$ automatically satisfy
the incompressibility condition in Eq. \eqref{eq:incompressibility_condition}.
It is important to bear in mind that this condition does not hold for a generic
antisymmetric tensor $\Lambda_{ij}$. Instead, it restricts the allowed generators to those
for which $\Lambda_{ij}$ is a closed two-form,
which has a clear physical interpretation. Volume-preserving diffeomorphisms are
the kinematic symmetry of an incompressible fluid, and only generators that preserve the
local volume element correspond to physical deformations. 
From the guiding-center perspective, the closedness of $\Lambda_{ij}$ reflects the higher-form
gauge redundancy of the tensorial coordinates $X^{ij}$, which encode oriented area degrees of
freedom rather than local shape deformations. Only closed $\Lambda_{ij}$ generate physical,
incompressible motions of the quantum Hall fluid.\\
We now show that the generators in Eq.\eqref{VPDgenerator} realize a closed Lie algebra
corresponding to volume-preserving diffeomorphisms. 
Let us consider two generators $G[\Lambda_{1}]$ and $G[\Lambda_{2}]$, such that the variation of
$G[\Lambda_{2}]$ under the transformation generated by $G[\Lambda_{1}]$ is given by
\begin{equation}
	\delta_{\Lambda_{1}}G[\Lambda_{2}]
	=
	\int d^{3}\mathbf{r}\,\delta_{\Lambda_{1}}\Lambda_{2\,ij}(\mathbf{r})\,X^{ij}(\mathbf{r}).
\end{equation}
By employing the transformation law in Eq.\eqref{VPDvariation},
\begin{equation}
	\delta_{\Lambda_{1}} x^{m}
	=
	\frac{1}{eH}\,\epsilon^{mkl}\Lambda_{1\,kl}(\mathbf{r}),
\end{equation}
the induced variation of the tensor field $\Lambda_{2\,ij}(\mathbf{r})$ is
\begin{equation}
	\delta_{\Lambda_{1}}\Lambda_{2\,ij}
	=
	\delta x^{m}\,\partial_{m}\Lambda_{2\,ij}
	=
	\frac{1}{eH}\,\epsilon^{mkl}\Lambda_{1\,kl}\,\partial_{m}\Lambda_{2\,ij}.
\end{equation}
Since the tensorial guiding-center coordinates themselves do not transform,
$\delta_{\Lambda_{1}}X^{ij}=0$, we obtain
\begin{equation}
	\delta_{\Lambda_{1}}G[\Lambda_{2}]
	=
	\int d^{3}\mathbf{r}\,
	\frac{1}{eH}\,\epsilon^{mkl}\Lambda_{1\,kl}\partial_{m}\Lambda_{2\,ij}\,X^{ij}.
\end{equation}
Antisymmetrizing under $1\leftrightarrow 2$, the Dirac bracket of the generators
takes the form
\begin{equation}
	\{G[\Lambda_{1}],G[\Lambda_{2}]\}_{D}
	=
	\delta_{\Lambda_{1}}G[\Lambda_{2}]
	-
	\delta_{\Lambda_{2}}G[\Lambda_{1}]
	=
	G[\Lambda_{3}],
	\label{VPDclosure}
\end{equation}
where the composite parameter $\Lambda_{3}$ is given by
\begin{equation}
	\Lambda_{3\,ij}
	=
	\frac{1}{eH}
	\left(
	\epsilon^{mkl}\Lambda_{1\,kl}\partial_{m}\Lambda_{2\,ij}
	-
	\epsilon^{mkl}\Lambda_{2\,kl}\partial_{m}\Lambda_{1\,ij}
	\right).
	\label{Lambda3}
\end{equation}
Eqs.\eqref{VPDclosure}--\eqref{Lambda3} define a closed algebra that
realizes the volume-preserving diffeomorphisms in tensorial form. The
nontrivial structure arises entirely from the geometric action of the generators
on spatially dependent tensor fields, while the underlying Dirac brackets remain
binary and associative.  Moreover, we emphasize that the resulting algebra should be understood as a representation
of volume-preserving deformations acting on the label-dependent guiding-center
fields, rather than as the geometric algebra of diffeomorphisms of physical space.\\
This construction should be contrasted with the standard Nambu-bracket formulation of
three-dimensional incompressible fluids, where infinitesimal transformations take the form
$\delta x^{i} \sim \{x^{i}, h_1, h_2\}$. While suggestive, Nambu brackets notoriously
lack a canonical associative operator representation \cite{Szabo}. In our present framework, the two scalar
functions $(h_1,h_2)$ could be seen as effectively packaged into a single antisymmetric tensor $\Lambda_{ij}$,
and the resulting algebra is realized entirely within ordinary associative mechanics.
Upon quantization, the generators $G[\Lambda]$ become operators whose commutators reproduce the
quantum version of the volume-preserving diffeomorphism algebra. The long-wavelength limit of
the three-dimensional GMP algebra derived in Section~2 is therefore identified with the quantum
algebra of incompressible deformations of a Hall fluid of quasi-strings.

\section{Topological Quantum Field Theory and DD invariant}

In this section, we discuss the topological aspects of the three-dimensional quantum Hall fluids of quasi-strings and their connection with the noncommutative geometry derived previously. In particular, we will show that the projected guiding-center algebra derived in Sec.~2 admits a consistent correspondence with a $(3+1)$-dimensional topological quantum field theory, namely the Abelian BF+BB theory \cite{Baez,Simon}.
Let us start considering Its corresponding metric-independent action, which is given by
\begin{equation}
	S_{BF}
	=
	\frac{\kappa_{DD}}{2 \pi}
	\int d^4x \,
	\epsilon^{\mu\nu\rho\sigma}
	B_{\mu\nu}\partial_\rho A_\sigma
	+
	\frac{\kappa_{DD}}{4 \pi}
	\int d^4x \,
	\epsilon^{\mu\nu\rho\sigma}
	B_{\mu\nu} B_{\rho\sigma},
\end{equation}
where $B$ is the Kalb-Ramond field and $A$ is a vector gauge field. Importantly, $\kappa_{DD}$ is the quantized level that coincides with the Dixmier-Douady Invariant (DD) \cite{Murray1996} via tensorial coordinates in the semiclassical regime \cite{Palumbo2025}. Moreover, the DD invariant is related to a tensor Berry connection and related bundle gerbe, which is nothing but a momentum-space Kalb-Ramond field. In this sense, our current work provides a dual real-space version of the gerbe structure in momentum-space discussed in Ref. \cite{Palumbo2025} in the context of three-dimensional Hall fluids of quasi-strings.\\
To derive the canonical quantization of the above action, we perform a $(3+1)$ decomposition and writing spatial indices $i,j,m=1,2,3$, the $BF$ term becomes
\begin{equation}
	S_{BF}
	=
	\frac{\kappa_{DD}}{2 \pi}
	\int dt\, d^3x \,
	\epsilon^{ijm}
	B_{ij}\, \partial_t A_m
	+ \text{(constraints)}.
\end{equation}
Although the $B B$ term contains no time derivatives and therefore does not affect the canonical structure, nevertheless it is essential in order to break time-reversal symmetry in agreement with the symmetry features of quantum Hall fluids.
The momentum conjugate to $A_m$ reads
\begin{equation}\label{canonicalBF}
	\Pi^m_A
	=
	\frac{\delta S_{BF}}{\delta(\partial_t A_m)}
	=
	\frac{\kappa_{DD}}{2 \pi} \epsilon^{mij} B_{ij},
\end{equation}
such that the canonical equal-time commutation relation is given by
\begin{equation}
	[A_m(\mathbf r), B_{ij}(\mathbf r')]
	=
	\frac{2\pi i}{\kappa_{DD}}
	\epsilon_{mij}
	\delta^{(3)}(\mathbf r-\mathbf r'),
	\label{canonical_BF}
\end{equation}
where we have replaces the expression for $\Pi^l_A$ from Eq. (\ref{canonicalBF}).
In order to compare the above expression with the microscopic guiding-center algebra, we must consider smeared operators corresponding to localized excitations \cite{Haag1996}. 
In fact, since $A_m(\mathbf r)$ and $B_{ij}(\mathbf r)$ are operator-valued distributions, physical observables are defined via smearing with smooth test functions. Localized excitations correspond to wavepackets, and the associated collective operators are obtained by projecting onto these modes.
Let $f(\mathbf r)$ and $g(\mathbf r)$ be normalized wavepackets and define
\begin{equation}
	\hat A^m
	=
	\int d^3r\, f(\mathbf r)\, A^m(\mathbf r),
	\qquad
	\hat B^{ij}
	=
	\int d^3r\, g(\mathbf r)\, B^{ij}(\mathbf r).
\end{equation}
By employing Eq. \eqref{canonical_BF}, we obtain
\begin{equation}
	[\hat A^m, \hat B^{ij}]
	=
	\frac{2\pi i \, V_{fg}}{\kappa_{DD}}
	\epsilon^{mij},
\end{equation}
where $V_{fg}$ is a volume element related to $f$ and $g$.
By choosing $f(\mathbf r)$ and $g(\mathbf r)$ such that
\begin{equation}
	\int d^3 r\, f(\mathbf r)\, g(\mathbf r) = 1,
\end{equation}
or, equivalently, by absorbing the overlap factor $V_{fg}$
into the normalization of the collective operators, we can express the smeared commutator in a simplified canonical form.
A direct comparison with the microscopic three-dimensional guiding-center algebra $	[x^m, X^{ij}]=i \ell_H^3 \epsilon^{mij}$, allow us to obtain the following identifications
\begin{equation}
	x^m \leftrightarrow \ell_H\, \sqrt{\frac{\kappa_{DD}}{2 \pi}} \, \hat {A}^m,
	\qquad
	X^{ij} \leftrightarrow \ell^2_H\, \sqrt{\frac{\kappa_{DD}}{2 \pi}} \, \hat{B}^{ij}.
\end{equation}
The above relations show that the BF+BB theory provides the dimensionless algebraic structure with the corresponding level associated to the DD invariant (topological sector), while the physical magnetic volume is fixed by microscopic input, i.e. the background flux density (geometric sector) and encoded through operator normalization.
This is the direct three-dimensional analogue of the well-known situation in the two-dimensional quantum Hall effect, where the guiding-center noncommutativity $[x^i,x^j]=i\ell_B^2 \epsilon^{ij}$ is related to the background magnetic field while the Chern-Simons theory \cite{Dunne2002,Palumbo2014} gives the algebraic structure with its quantized level related to the first Chern number.\\
It is instructive to compare the present construction with the hydrodynamic
BF description of three-dimensional incompressible polar fluids developed
in Ref.~\cite{Tiwari}. In that work, an extended
GMP-type algebra arises from the canonical quantization of a $(3+1)$-dimensional
BF theory describing coupled particle and vortex densities. The long-wavelength
limit realizes the algebra of VPDs,
providing an effective hydrodynamic description of incompressibility. In Ref.~\cite{Tiwari}, the (equal-time) commutation relation in the BF theory is taken with respect to the dual field of $B_{ij}$, namely $B^k=(1/2) \epsilon^{kij}B_{ij}$, such that
\begin{equation}
	[A^m(\mathbf r), B^{k}(\mathbf r')] \propto i\, \delta^{mk}
	\delta^{(3)}(\mathbf r-\mathbf r'), \hspace{0,3cm} B^k= \frac{1}{2}\epsilon^{kij}B_{ij}.
	\label{canonical_BF2}
\end{equation}
In contrast, our construction starts from a microscopic strong-field $H$
projection of extended quasi-string degrees of freedom.
However, a precise connection with the above BF quantization structure becomes manifest in our case
upon (Hodge) dualizing the antisymmetric tensor $X^{ij}$. In three spatial
dimensions we define the dual axial-vector variables
\begin{equation}
	\widetilde{X}^{k}
	\equiv
	\frac{1}{2}\,\varepsilon^{kij} X_{ij},
	\label{eq:hodge_def}
\end{equation}
which measure the oriented area element perpendicular to direction $k$.
By using the identity $\varepsilon^{mij}\varepsilon_{kij}= 2\,\delta^{m}_{k},$
we obtain
\begin{equation}
	[x^{m},\widetilde{X}^{k}]
	=
	i\ell_H^{3}\delta^{mk}.
	\label{eq:canonical_form}
\end{equation}
Therefore, in the dual formulation, our algebra reduces to a canonical
Heisenberg–Weyl algebra and can be directly related to the canonical
quantization of the BF theory in Eq.~(\ref{canonical_BF2}).
It is important to stress, however, that Eq.~(\ref{eq:canonical_form})
describes the algebra of the \emph{projected} (guiding-center) theory.
In the strong-field limit, the microscopic kinetic momentum is eliminated
by Dirac reduction, analogously to the lowest-Landau-level projection in
two dimensions. The reduced phase space is then spanned solely by
$(x^{m},X^{ij})$, and the dual variable $\widetilde{X}^{k}$ becomes the
only operator canonically conjugate to $x^{m}$. Accordingly,
\begin{equation}
P_{eff}^k\equiv \widetilde{X}^{k}/\ell_H^{3}
\end{equation}
acts as an effective translation generator
in the projected Hilbert space: momentum is not an independent microscopic
degree of freedom but emerges geometrically from the area variables
induced by the background three-form field.
Thus, in the three-dimensional Hall fluids of quasi-strings, spatial motion is intrinsically tied to geometric area fluctuations, namely translating the center of mass necessarily
reconfigures the internal flux carried by the excitation. In this way,
transport properties emerge from extended-object kinematics rather than
from conventional particle momentum.
From this perspective, the hydrodynamic density–vortex algebra of
Ref.~\cite{Tiwari} can be viewed as an effective dual representation of
the extended-object guiding-center kinematics derived here. Although both
approaches lead to a similar long-wavelength noncommutative structure,
their starting points and physical interpretation are fundamentally different. In contrast to Ref.~\cite{Tiwari}, our noncommutative geometry arises prior to any
coarse-graining, directly from the strong-field projection of a
microscopic generalized phase space, consistently with the semiclassical
formalism introduced in Ref.~\cite{Palumbo2025}. The tensorial guiding-center
coordinates $X^{ij}$ follow from the intrinsic coupling of quasi-strings
to a background three-form field and encode genuine geometric degrees
of freedom.\\
This microscopic derivation parallels the role of the magnetic length in
two dimensions: just as $\ell_B$ controls the guiding-center algebra in
the quantum Hall effect, the magnetic length $\ell_H$ fixes the
three-dimensional noncommutative structure and its associated GMP algebra.
The BF+BB theory and its quantized level $\kappa_{\rm DD}$ then emerge as
the compatible topological sector rather than as a starting assumption.
Hence, our framework provides a geometric and microscopic origin for the
three-dimensional GMP algebra and recovers the hydrodynamic BF description
as its effective long-wavelength limit.

\section{Conclusions and outlook}

In this work we have developed a continuum framework for three-dimensional quantum
Hall physics based on quantum fluids of extended objects in the presence of a strong
background three-form field $H=dB$. Treating the strong-field regime via Dirac
brackets and projecting onto $\sigma$-integrated observables, we identified a set
of guiding-center variables consisting of vectorial coordinates $x^{m}$ and
antisymmetric tensorial coordinates $X^{ij}$. We showed that these variables obey a
closed, associative noncommutative algebra, providing a three-dimensional
generalization of the GMP algebra appropriate to extended
objects.
The background three-form $H$, which defines a two-plectic structure \cite{Rogers,Saemann2012} at the
microscopic level, plays a central role in generating the reduced guiding-center
algebra. 
While formally analogous to the lowest-Landau-level construction in two-dimensional
quantum Hall systems, the present framework is distinguished by the appearance of
tensorial guiding-center degrees of freedom reflecting the extended nature of the
underlying excitations.
Our results establish an analogy between two- and three-dimensional
quantum Hall physics: in both cases, strong background fields freeze fast
degrees of freedom, leading to noncommutative guiding-center coordinates and
GMP-type algebras that encode incompressibility and collective dynamics.
In three dimensions, some of the relevant kinematic variables are tensorial and reflect
the coupling of the system to higher-form gauge fields and extended excitations (quasi-strings). 
Finally, we have shown a correspondence between the canonical commutators of a topological quantum field theory and the microscopic guiding-center algebra, establishing a precise bridge between the topological bulk description and the noncommutative kinematics of the quantum Hall fluid. 
All the main aspects analyzed in this work as summarized in the Table 1 below.
The present work is intentionally kinematic in nature, while dynamical realizations and microscopic lattice models are left for future investigations.
For instance, a clear and precise understanding of a possible charge fractionalization carried by quasi-strings in three dimensions beyond an effective topological field theory description (see for instance Refs \cite{Tizzano,Palumbo2022} about fractional quantum Hall effect for membranes) would be high desirable.
Moreover, an important future direction would be the identification of three-dimensional GMP algebra and its geometric consequences in lattice systems and band models.
In particular, it would be highly desirable to realize this algebra 
in (interacting) three-dimensional topological insulators in symmetry class AIII, where nearly flat topological bands can be characterized by the DD invariant and
tensor Berry connections associated to momentum-space bundle gerbes \cite{Palumbo2019, Zhu2020, Palumbo21, Jankowski2025}, along this direction, it has been already shown the emergence of a BF theory in interacting topological insulators \cite{Cirio} although the charge fractionalization was not analyzed.
Such interacting lattice systems provide a promising platform to emulate the physics of extended
objects discussed in this work by using purely electronic degrees of freedom.
In this context, the three-dimensional GMP algebra may control density
correlations and propagating collective modes in
topological bands, in close analogy with the role played by the GMP algebra in
fractional Chern insulators \cite{Roy2014}. 
More broadly, the framework developed here opens new avenues for exploring
higher-dimensional incompressible quantum fluids, higher-form gauge responses,
and the interplay between topology and noncommutative geometry.
We expect that further connections with band topology, interacting topological
phases, and engineered quantum systems will continue to shed light on the
physical realization and consequences of higher-form quantum Hall phenomena.
\begin{table}[h]
	\centering
	\caption{Comparison between two-dimensional and three-dimensional quantum Hall fluids (QHFs)}
	\label{tab:2d3d_comparison}
	\begin{tabular}{l|c|c}
		\hline
		\textbf{Aspects} & \textbf{2D QHF} & \textbf{3D QHF} \\
		\hline
		Background field & $F$ (2-form) & $H$ (3-form) \\
		\hline
		Fundamental objects & Point particles & Extended objects (strings) \\
		\hline
		Guiding centers & $x^m$ & $(x^m, X^{ij})$ \\
		\hline
		Characteristic scale & $\ell_B = 1/\sqrt{e\mathcal{B}}$ & $\ell_H = 1/\sqrt[3]{eH}$ \\
		\hline
		Noncommutative geometry & $[x^i, x^j] = i\ell_B^2 \epsilon^{ij}$ & $[x^m, X^{ij}] = i\ell^3_H \epsilon^{mij}$ \\
		\hline
		Sine/GMP algebra & $-2i\sin\left(\frac{\ell_B^2}{2} q \wedge q'\right)$ & $-2i\sin\left(\frac{\ell_H^3}{2} q \wedge Q\right)$ \\
		\hline
		Diffeomorphisms & Area-preserving diffs & Volume-preserving diffs \\
		\hline
		Geometric structure & Symplectic & Two-plectic  \\
		\hline
		Topological action & Chern-Simons & BF+BB  \\
		\hline
		Topological invariant & Chern number & DD number\\
		\hline
	\end{tabular}
\end{table}

\section{Acknowledgments}

We thank Wojciech J. Jankowski for useful discussions and comments.

\section{Appendix A}
\label{app:transverse-inverse}

For a generic reparametrization-invariant observable
$\mathcal O=\oint d\sigma\,\mathcal O[X(\sigma)]$, terms proportional to
$\partial_\sigma\delta(\sigma-\sigma')$ contribute as
\begin{equation}
	\oint d\sigma\, B_{ij}(X(\sigma))\,\partial_\sigma\delta(\sigma-\sigma')
	=
	-\oint d\sigma\, \partial_\sigma B_{ij}(X(\sigma))\,\delta(\sigma-\sigma') ,
\end{equation}
where integration by parts produces no boundary term for a closed string.
For a uniform background $H=dB$ and a linear gauge choice for $B_{ij}$,
$\partial_\sigma B_{ij}(X(\sigma))\propto H_{ijk}\partial_\sigma X^k$,
which does not contribute to Dirac brackets of $\sigma$-integrated observables.\\

In the uniform strong-field (projected) regime of Sec.~2, the leading contribution to the constraint kernel
can be written in the local form
\begin{equation}
	C_{ij}(\sigma,\sigma')
	\;=\;
	-eH\,\epsilon_{ijk}\,\partial_{\sigma}X^{k}(\sigma)\,\delta(\sigma-\sigma')\,,
	\label{eq:Ctilde_app}
\end{equation}
At each value of $\sigma$, the $3\times 3$ matrix $C_{ij}(\sigma)$ has a null vector along the
tangent direction $\partial_{\sigma}X^{i}(\sigma)$:
\begin{equation}
	C_{ij}(\sigma)\,X^{\prime j}(\sigma)
	=
	-eH\,\epsilon_{ijk}\,X^{\prime k}(\sigma)\,X^{\prime j}(\sigma)
	=0.
	\label{eq:nullvector_app}
\end{equation}
Hence $C$ is rank-$2$ and does not admit a genuine inverse on the full three-component
space. This degeneracy is expected and reflects the redundancy associated with $\sigma$-reparametrizations:
infinitesimal variations $\delta X^i(\sigma)\propto \partial_{\sigma}X^{ i}(\sigma)$ correspond to relabeling the
worldsheet coordinate rather than to a physical deformation of the embedded loop.
To define an inverse on the physical subspace, introduce the unit tangent and the transverse projector
\begin{equation}
	t^i(\sigma)\equiv \frac{\partial_{\sigma}X^{ i}(\sigma)}{|\partial_{\sigma}X(\sigma)|},
	\qquad
	(P_\perp)^i{}_j(\sigma)\equiv \delta^i{}_j - t^i(\sigma)\,t_j(\sigma),
	\label{eq:projector_app}
\end{equation}
which projects onto the plane orthogonal to $\partial_{\sigma}X^{ i}(\sigma)$.
A convenient \emph{transverse inverse} (Moore--Penrose-type pseudoinverse) for $C$ is the
local distributional kernel
\begin{equation}
	C^{-1}_{ij}(\sigma,\sigma')
	\equiv
	\frac{1}{eH}\,\epsilon_{ijk}\,\frac{\partial_{\sigma}X^{ k}(\sigma)}{|\partial_{\sigma}X(\sigma)|^{2}}\;\delta(\sigma-\sigma').
	\label{eq:Cinv_transverse_app}
\end{equation}
Its defining property is that it inverts $C$ on the transverse subspace, i.e. it reproduces the
projector $P_\perp$ rather than the full identity
\begin{align}
	\int d\sigma''\;
	C_{ik}(\sigma,\sigma'')\, C^{-1}_{kj}(\sigma'',\sigma')
	&=
	\left[-eH\,\epsilon_{ik\ell}\partial_{\sigma}X^{ \ell}(\sigma)\right]
	\left[\frac{1}{eH}\,\epsilon_{kjm}\,\frac{\partial_{\sigma}X^{ m}(\sigma)}{|\partial_{\sigma}X(\sigma)|^2}\right]\delta(\sigma-\sigma')
	\nonumber\\
	&=
	-\epsilon_{ik\ell}\epsilon_{kjm}\,
	\frac{\partial_{\sigma}X^{ \ell}\partial_{\sigma}X^{ m}}{|\partial_{\sigma}X|^2}\,\delta(\sigma-\sigma')
	\nonumber\\
	&=
	\left(\delta_{ij}-\frac{\partial_{\sigma}X_i \partial_{\sigma}X_j}{|\partial_{\sigma}X|^2}\right)\delta(\sigma-\sigma')
	=
	-(P_\perp)_{ij}(\sigma)\,\delta(\sigma-\sigma').
	\label{eq:CCinv_projector_app}
\end{align}
In the last step we used the standard identity
$\epsilon_{ik\ell}\epsilon_{kjm}=\delta_{ij}\delta_{\ell m}-\delta_{im}\delta_{\ell j}$.
Eq.~\eqref{eq:CCinv_projector_app} shows that $C^{-1}$ coincides with the usual inverse
on the subspace orthogonal to $\partial_{\sigma}X$, while it annihilates the tangential (null) direction.
Consequently, Dirac brackets computed with $C^{-1}$ are unambiguous for
reparametrization-invariant observables, whereas quantities sensitive to the tangential sector require a
choice of worldsheet gauge (equivalently, a completion of the pseudoinverse of the transverse
subspace).\\

For completeness, we explicitly verify that the remaining projected Dirac brackets
among the guiding-center variables vanish identically.
We first consider $\{x^{i},x^{j}\}_{D}$. The canonical Poisson bracket vanishes,
$\{x^{i},x^{j}\}=0$, since $x^{i}$ is linear in $X^{i}(\sigma)$. The Dirac correction
term involves
\begin{equation}
	\{x^{i},\phi_{k}(\sigma)\}=\frac{1}{2\pi}\delta^{i}_{k},
\end{equation}
and the inverse constraint matrix $C^{-1}_{ij}(\sigma,\sigma')$.
Using the explicit expression for $C^{-1}$, one finds
\begin{equation}
	\{x^{i},x^{j}\}_{D}
	=
	-\frac{1}{(2\pi)^2}
	\frac{1}{eH}
	\epsilon_{ijk}
	\oint d\sigma\,
	\frac{\partial_\sigma X^{k}(\sigma)}{|\partial_\sigma X(\sigma)|^2}.
\end{equation}
Since the string is closed, $\oint d\sigma\,\partial_\sigma X^{k}=0$, and therefore
\begin{equation}
	\{x^{i},x^{j}\}_{D}=0 .
\end{equation}
We next evaluate $\{X^{ij},X^{kl}\}_{D}$. The canonical bracket vanishes identically,
$\{X^{ij},X^{kl}\}=0$. The relevant brackets with the constraints are
\begin{equation}
	\{\phi_{m}(\sigma),X^{ij}\}
	=
	\delta^{i}_{m}\partial_\sigma X^{j}
	-
	\delta^{j}_{m}\partial_\sigma X^{i}.
\end{equation}
Substituting into the Dirac-bracket definition yields
\begin{align}
	\{X^{ij},X^{kl}\}_{D}
	&=
	\frac{1}{eH}
	\oint d\sigma\,
	\epsilon_{mnp}
	\frac{\partial_\sigma X^{p}}{|\partial_\sigma X|^2}
	(\delta^{i}_{m}\partial_\sigma X^{j}-\delta^{j}_{m}\partial_\sigma X^{i})
	(\delta^{k}_{n}\partial_\sigma X^{l}-\delta^{l}_{n}\partial_\sigma X^{k}) .
\end{align}
Each term contains a totally antisymmetric contraction of three identical tangent
vectors $\partial_\sigma X$, which vanishes identically. Hence,
\begin{equation}
	\{X^{ij},X^{kl}\}_{D}=0 .
\end{equation}
These results confirm that the only nonvanishing projected Dirac brackets are the
mixed brackets $\{x^{m},X^{ij}\}_{D}$.

\section{Appendix B}
In this Appendix we summarize the algebraic conventions and operator
definitions underlying the dual guiding-center formulation used in the
main text.
We work on a three-dimensional flat space and Latin indices $i,j,k,\dots$ run over spatial components $\{x,y,z\}$.  We use the Levi-Civita symbol $\varepsilon^{ijk}$ with $\varepsilon^{123}=+1$. The basic operators are given by
\begin{itemize}
	\item Vector coordinates $x^i$ and their scalar projected densities
	\begin{equation}\label{rho_def}
		\rho(\mathbf{q}) := e^{\,i q_m x^m},\qquad q_m\in\mathbb{R}.
	\end{equation}
	\item Antisymmetric tensor coordinates $X^{ij}=-X^{ji}$ and tensor projected densities
	\begin{equation}\label{T_def}
		\Upsilon(\mathbf{Q}) := e^{\,i Q_{ij} X^{ij}},\qquad Q_{ij}=-Q_{ji}.
	\end{equation}
\end{itemize}
The goal of this Appendix is to compute the commutators $[\rho(q),\Upsilon(Q)]$,
starting from the following noncommutative geometry
\begin{align}
	&[x^i,x^j] = [X^{ij},X^{kl}] = 0, \nonumber \\
	&[x^m,X^{ij}] = i\,\ell^3_H\,\varepsilon^{mij}. \label{xX_central}
\end{align}
The last relation defines a dimensional volume $\ell_H^3$. Note that Eq.\eqref{xX_central} is taken to be \textit{central} (a c-number) so it commutes with both $x$ and $X$, enabling the Baker--Campbell--Hausdorff (BCH) simplifications below.
Let us now define the two generator operators
\begin{equation}
	A := i\,q_m x^m,\qquad B := i\,Q_{ij} X^{ij}.
\end{equation}
We can then compute the commutator of generators in Eq.\eqref{xX_central}
\begin{align}
	[A,B] &= [i q_m x^m,\; i Q_{ij} X^{ij}] \nonumber \\
	&= - q_m Q_{ij} [x^m,X^{ij}] \nonumber \\
	&= - q_m Q_{ij} (i\ell^3_H\varepsilon^{mij}).
	\label{AB_comm}
\end{align}
The corresponding BCH identity is given by
\begin{equation}
	e^{A}e^{B} = e^{A+B+\tfrac12[A,B]},\qquad e^{B}e^{A} = e^{A+B-\tfrac12[A,B]},
\end{equation}
such that
\begin{align}
	[e^{A},e^{B}] &= e^{A+B}\Big(e^{\tfrac12[A,B]} - e^{-\tfrac12[A,B]}\Big) \nonumber \\
	&= 2\,e^{A+B}\sinh\!\Big(\tfrac12[A,B]\Big).\label{comm_sinh}
\end{align}
By substituting Eq.\eqref{AB_comm}, we obtain
\begin{align}
	\sinh\!\Big(\tfrac12[A,B]\Big) &= \sinh\!\Big(-\tfrac{i\ell_H^3}{2}q_m Q_{ij}\,\varepsilon^{mij}\Big) \nonumber \\ 
	&= -i\sin\!\Big(\tfrac{\ell_H^3}{2}q_m Q_{ij}\,\varepsilon^{mij}\Big).
\end{align}
Therefore the exact closed-form commutator reads
\begin{equation}\label{rhoT_commutator_exact}
	[\rho(\mathbf{q}),\Upsilon(\mathbf{Q})] = -2i\;\sin\!\Big(\tfrac{\ell^3_H}{2}\,q_m Q_{ij}\varepsilon^{mij}\Big)\; e^{\,i q_m x^m + i Q_{ij} X^{ij}}.
\end{equation}
For small arguments ($\ell_H^3 |q||Q|\ll1$) we can expand the sine to linear order and get
\begin{align}
	[\rho(\mathbf{q}),\Upsilon(\mathbf{Q})] \approx 
	-i\ell^3_H\, q_m Q_{ij}\varepsilon^{mij}\;e^{\,i q_m x^m + i Q_{ij} X^{ij}} + ....\label{rhoT_linear}
\end{align}
The leading term in this expansion is the analogue in three dimensions of the linearized GMP term in the two-dimensional case. In particular, the structure
$q_m Q_{ij}\varepsilon^{mij}$ reflects the underlying volume-preserving
diffeomorphism algebra generated by the guiding-center variables and is
controlled by the magnetic volume $\ell_H^3$. This confirms that the
noncommutative geometry derived in the main text consistently reproduces
the expected long-wavelength limit of the three-dimensional GMP algebra.

\newpage
\bibliographystyle{apsrev4-2}
\bibliography{references}

\end{document}